\documentclass[11pt, spanish]{article}
\usepackage{cite}
\usepackage{amsmath,amsfonts,amssymb}
\usepackage[small,bf,hang]{caption}
\usepackage{slashed}
\usepackage{color}
\usepackage{physics}

\usepackage{extarrows}
\usepackage{hyperref}

\usepackage{ulem}

\usepackage{geometry}
\def\hybrid{
        \topmargin 0pt
        \oddsidemargin 0pt
        \headheight 0pt \headsep 0pt
        \textwidth 6.25in 
        \textheight 9.5in 
        \marginparwidth .875in
        \parskip 5pt plus 1pt \jot = 1.5ex}

\hybrid

 \csname
@addtoreset\endcsname{equation}{section}

\usepackage{epsfig}
\usepackage{color}
\usepackage{graphics}
\usepackage{graphicx}
\usepackage{slashed}

\usepackage{amsthm}
\usepackage{amssymb}
\usepackage{amsmath}
\usepackage{amsfonts}

\numberwithin{equation}{section}

\newcommand{\be}{\begin{equation}}
\newcommand{\ee}{\end{equation}}
\newcommand{\ba}{\begin{eqnarray}}
\newcommand{\ea}{\end{eqnarray}}

\input xy
\xyoption{all}

\begin{document}
\begin{titlepage}
\rightline{}
\begin{center}
\vskip 1.5cm
 {\Large \bf{$\beta$ symmetry in type II Supergravities}}
\vskip 1.7cm

{\large\bf {Walter H. Baron$^1$, Nahuel A. Yazbek$^{2}$}}
\vskip 1cm

$^1$ {\it  Instituto de F\'isica de La Plata}, (CONICET-UNLP)\\
{\it Departamento de Matem\'atica, Universidad Nacional de La Plata. }\\
{\it C. 115 s/n, B1900, La Plata, Argentina}\\
wbaron@fisica.unlp.edu.ar\\
 
\vskip .3cm

$^2$ {\it   Instituto de Matemática Luis Santaló}, (CONICET-UBA)\\
{\it Ciudad Universitaria, Pabell\'on I, CABA, C1428ZAA, Argentina}\\
nyazbek@dm.uba.ar, 

\vskip .3cm

\vskip .1cm

\vskip .4cm

\vskip .4cm

\end{center}
\bigskip\bigskip

\begin{center} 
\textbf{Abstract}

\end{center} 
\begin{quote}  
A non geometric sector of the duality group emerging in Kaluza-Klein reductions is realized as an effective symmetry in the low energy action of uncompactified type II theories. This is achieved by extending the so called $\beta$ symmetry of the universal NS-NS sector to the R-R sector of type IIA, IIB and massive type IIA.

\end{quote} 
\vfill
\setcounter{footnote}{0}
\end{titlepage}


\section{Introduction}

Symmetries and dualities in String theory play a fundamental role. At the perturbative level, the very definition of the theory, including the field content, the interactions and even the space-time dimension is determined by symmetry arguments.  

The same holds by transitivity in the effective actions, at low energy. The symmetries inherited from string theory: diffeomorphisms, Lorentz, gauge and supersymmetry impose strong constraints on the allowed interactions at the supergravity level. However, these symmetries alone are not sufficient to fully fix them. Fortunately, in addition to symmetries, dualities come into play. Dualities establish connections between seemingly different quantum field theories in backgrounds with opposite features ($e.g.$, large/small compact spaces) or different regimes (for instance, at strong/weak couplings). 

This article deals with T-duality, whose origin can be traced back to the extended nature of the strings. At quantum level, it is related to a symmetry in the string spectrum that involves exchanging Kaluza-Klein (KK) modes and winding modes around closed directions in d-dimensional toroidal compactifications. This symmetry is described by the action of a discrete $O(d,d,{\mathbb{Z}})$ group, but it is enhanced into $O(d,d,{\mathbb{R}})$ at low energy, when all massive fields are truncated.

This symmetry was explicitly realized at low energy, in the two-derivative effective actions of string theory, by Meissner and Veneziano in \cite{MeissnerVeneziano} on 1-dimensional (cosmological) reductions. Subsequently, it was further extended to arbitrary dimensions by Maharana and Schwarz in \cite{Maharana:1992my} by considering general KK reductions. Interestingly, although $O(d,d)$ manifests itself in supergravity only after toroidal compactifications, the fact that this symmetry emerges after a proper reduction already constrains the allowed couplings in the parent higher dimensional theory in an indirect way. 

This simple observation has profound implications, specially after the work of Sen \cite{Sen:1991zi}, where closed string field theory arguments were used to promote a tree level $O(d)\times O(d)$ symmetry\footnote{Actually, this $O(d)\times O(d)$ is further completed into $O(d,d)$ when it is combined with $GL(d)$ and $b$-shifts.} emerging after KK to all orders in derivatives.  

Besides these arguments, the explicit verification of this symmetry beyond two-derivatives is not a straightforward task and it took some time until an explicit verification was achieved \cite{Meissner:1996sa}. The complexity arises from the fact that while both T-duality and Lorentz are symmetries of the compactified theory, they cannot be simultaneously made manifest. An explicit realization of $O(d,d)$ requires Lorentz non covariant field redefinitions in the internal space.

An alternative approach to address T-duality is given by Double Field Theory (DFT), where the $O(d,d)$ is realized already before compactification at the cost of introducing extra coordinates, dual to winding modes \cite{dft}. This is an interesting approach because the constraints on the effective action arising from T-duality are more easily implemented. Some interesting applications concern the uplifting of non-geometric gaugings and the derivation of general formulae for non linear uplift ansatze on gauged supergravities, leading to consistent truncations in lower dimensions, both followings from the so called generalized Scherk Schwarz compactifications \cite{GSS}.

Other relevant applications are the determination of $\alpha'$-corrections in the low energy limit, where an iterative method was developed \cite{GBdR} leading to an infinite tower of couplings (all order in derivatives) of the so called biparametric deformation of DFT \cite{Marques:2015vua}. Despite some interesting consistency checks \cite{Hronek:2021nqk} showing agreement with the bosonic and heterotic supergravities at sixth order in derivatives, it is known that extra couplings, not captured by this method, appear at eighth order in derivative and higher ($e.g.$ those corresponding to type II supergravities). Actually, it is not even clear so far if such missing interactions admit an embedding in a T-dual formulation, like DFT \cite{Hronek:2020xxi}. 

Recently, a different approach was introduced in \cite{Baron:2022but}, \cite{Baron:2023qkx} where the non geometric sector of the T-duality group is promoted as an effective symmetry in the canonical formulation of supergravity, avoiding the standard procedure of the KK reduction or the introduction of extra coordinates and so offering an alternative method, free of the obstructions raised in \cite{Hronek:2020xxi} and provides a new perspective on incorporating the non-geometric aspects of T-duality into the framework of supergravity. 

This program focuses on infinitesimal duality transformations,\footnote{We are therefore not considering the so called Buscher rules. For instance, the consequences of the discrete $O(1,1,{\mathbb Z})$ symmetry on string theory due to circular compactifications is therefore not covered. See \cite{Garousi} for some recent progress on $\alpha'$-corrections, after exploiting this symmetry.} so it deals with the connected component of $O(d,d)$, containing the identity element of the group. This is organized in two sectors dubbed geometric and non geometric, respectively. The former is generated by $GL(d)$ and  ${\mathbb R}^{d(d-1)/2}$ (referred to as b-shifts), which are already symmetries of the higher dimensional theory, while the latter is a symmetry only after compactification. Notwithstanding, it was shown in \cite{Baron:2022but},\cite{Baron:2023qkx}  that also the non-geometric sector,  parameterized by a constant bivector $\beta^{\mu\nu}$, can be considered as an {\it effective symmetry} of the universal NS-NS Lagrangian prior to any compactification (the invariance of the action was already established in \cite{Andriot:2014uda} within the context of $\beta$-supergravity. See comments below), with the following transformation rules 
\begin{eqnarray}
\delta_{\beta} e_{\mu}{}^{a} &=& - e_{\mu}{}^{b} b_{bc} \beta^{ca}  \;, \cr 
\delta_{\beta} b_{\mu\nu} &=& - \beta_{\mu\nu} - b_{\mu\rho} \beta^{\rho\sigma} b_{\sigma\nu}  \;, \cr 
\delta_{\beta} \phi &=& \frac12 \beta^{\mu\nu} b_{\mu\nu}  \;,\label{NSbeta}
\end{eqnarray}
when these are supplemented with the constraint 
\begin{eqnarray}
\beta^{\mu\nu}\partial_{\nu}=0.    \label{betaConstraint}
\end{eqnarray}     

Interestingly, while diffeomorphisms, Lorentz and gauge symmetry constrain the allowed interactions up to an arbitrary function of the dilaton and some undetermined coefficients 
\begin{eqnarray}
{\cal L}_{f,c,d,e}= f(\phi) \left(R + c\Box \phi + d(\nabla\phi)^2 + e H^2\right)\;,     
\end{eqnarray}
the requirement of beta symmetry fully fixes the dynamics at two derivative order
\begin{eqnarray}
0 &=& \delta_\beta[\sqrt{-g} {\cal L}_{f,c,d,e}] \cr
&=& \sqrt{-g}f(\phi)\left[(c-4)\nabla^{a}\left(\beta^{bc}H_{abc}\right)-(12 e + 1) \nabla^{a} \beta^{bc} H_{abc} + 2(c+d) \nabla^{d}\phi \beta^{bc} H_{abc}\right]\cr
&+& \frac12 \sqrt{-g} \beta^{ab} b_{ab}\left(2f(\phi) + f'(\phi)\right)\left[R + c\Box \phi + d(\nabla\phi)^2 + e H^2\right] \;, \label{dLNS}
\end{eqnarray}
leading to the NS-NS universal Lagrangian of string supergravity 
\begin{eqnarray}
{\cal L}_{NS}= e^{-2\phi} \left(R + 4\Box \phi - 4 (\nabla\phi)^2 -\frac{1}{12} H^2\right)\;.     \label{LNS}
\end{eqnarray}

This Lagrangian admits a two-parameter family of deformations consistent with string dualities at fourth derivative order. Specific choices in the parameter space correspond to the bosonic and heterotic supergravities\footnote{The trivial case is identified with the type II supergravities, whose first corrections appear at eighth order in derivatives.}. In \cite{Baron:2022but} the $\alpha'$-corrections of the transformation rules (\ref{NSbeta}) were obtained for the whole family of deformations by requiring invariance of the Lagrangian. An independent derivation (from DFT) was presented in \cite{Baron:2023qkx}. 

The $O(d,d)$ sector described by $\beta$ was also discussed in a related context in \cite{Andriot:2012wx, Andriot:2013xca, Andriot:2014uda}, within the so called $\beta$-supergravity scheme. There, the manifest symmetries differ from the standard formulation of supergravity and the $\beta$ symmetry becomes {\it geometric}, while the b-shifts are interpreted as {\it non-geometric}. The 2-derivative Lagrangian of the NS-NS sector in the standard metric scheme, in the form presented in \cite{Andriot:2014uda},\footnote{It differs from (\ref{LNS}) by a total derivative.} and the same Lagrangian in the $\beta$-supergravity scheme differ by a total derivative and field redefinitions. This is, as the authors of $loc.cit.$ properly stated (see section 4.1.2 in the aforementioned reference), an indirect proof that the 2-derivative NS-NS Lagrangian of the standard metric formulation should be invariant, at least up to a total derivative, under $\beta$-transformations.

The non-geometric sector of the T-duality transformations admits the interpretation of an Abelian Yang Baxter (YB) deformation (see for instance \cite{YBOdd}). In this approach, these transformations are implemented as a solution generating technique and the $\beta$ supergravity mentioned above plays a crucial role in the connection between generalized supergravity $e.o.m.$ and Classical Yang Baxter equations\cite{Bakhmatov:2018bvp}. See \cite{YBRR} for related works considering the RR sector. YB deformations admit generalizations relating it to Non Abelian T-dualities and Poisson Lie T-duality and have the nice property that can be used to generate integrable deformations (see for instance \cite{PLTDRR}).  

Another interesting method that achieves $O(d)\times O(d)$ symmetry in supergravity in 10 dimensions is discussed in reference \cite{Hyakutake:2023oiy}. In this approach, duality transformations of the dimensionally reduced fields are used to formally promote this symmetry to higher dimensions. This extension enables the determination of interactions in the uncompactified theory by construction of $O(d)\times O(d)$ invariant blocks.

In this article we will extend the $\beta$ symmetry to the R-R sector at two derivative order. This extension will be carried out within the democratic formulation of type II theories, which we will review in section 2, in order to set the notation and to ensure a self-contained discussion. In section 3 we analyze Kaluza Klein compactifications and exhibit how duality acts on $GL(d)$ multiplets. Moving on to section 4, we will utilize the previously derived expressions to establish the non-geometric sector of the duality group as an effective symmetry of the R-R sector in 10 dimensions. We will present a detailed verification of the invariance of the Lagrangian ${\cal L}_{R}$ for type IIB and IIA. In addition, we also display the extension to the massive type IIA case. Compatibility with duality relations of the democratic formulation, closure of the symmetry algebra and consistency with equations of motion ($e.o.m.$) are also discussed. Finally, we present the conclusions in section 5.

\section{Democratic formulation of type II theories}

The natural framework to study the $O(d,d)$ symmetry enhancement on KK reductions of type II theories is the so called democratic formulation \cite{Bergshoeff:2001pv}. This approach is preferred because it naturally accommodates R-R potentials in an $O(d,d)$ multiplet, allowing for a linear action of the duality group (see $e.g.$ \cite{Fukuma:1999jt}). Additionally, a democratic like version of DFT was used to incorporate R-R potentials within the Double Field Theory framework \cite{DFTtypeII}-\cite{Catal-Ozer:2017ycb}. Therefore, it is reasonable to explore the extension of $\beta$ symmetry to the R-R sector in such a scheme.

In the standard formulation of type II supergravities (see for instance \cite{polcho}), p-form potentials are present for $p$ ranging from 0 to 4, with even (odd) $p$ for type IIB (IIA).  The Lagrangian includes kinetic terms $|F^{(p+1)}|^2$ as well as Chern-Simons interactions. In the case of IIB, the curvature $F^{(5)}$ is constrained to be self-dual, a condition that must be imposed only after obtaining the equations of motion. The type IIA case admits a further deformation, known as massive type IIA, which includes a non-dynamical curvature $F^{(0)}$. It is worth noting that in $D=10$ the theory actually allows for the inclusion of higher p-forms: $F^{(6)}$ to $F^{(10)}$ and indeed these fields are included in the theory, but as Hodge duals of the former ones.

In the democratic formulation\cite{Bergshoeff:2001pv}, in contrast, the Lagrangian contains quadratic Kinetic terms but not Chern-Simons. The field content now includes all allowed p-form in $D=10$, 
\begin{eqnarray}
{\cal S}_R=\int d^{10}x \sqrt{-g}\; {\cal L}_{R} = -\frac14 \int d^{10}x \sqrt{-g}\; \sum_{n} |F^{(n)}|^2  = \frac14 \int\;  \sum_{n} F^{(n)}\wedge *F^{(n)}\;, \label{LR} 
\end{eqnarray}
where $n$ runs over $1,3,5,7,9$ for type IIB and $2,4,6,8$ for type IIA, respectively. Here we used standard notation $\displaystyle{F^{(n)}=\frac{1}{n!}F^{(n)}_{\mu_1 ... \mu_n}\, dx^{\mu_1}\wedge ... \wedge dx^{\mu_n}}$ and $\displaystyle{|F^{(n)}|^2=\frac{1}{n!}F^{(n)}_{\mu_1 ... \mu_n} F^{(n)}{}^{\mu_1 ... \mu_n}}$.

The field strengths are defined in a very compact way by means of formal sums of forms\footnote{Being $e^{-b}$ a formal sum of even forms, it then commutes with respect to the wedge product: $e^{-b}\wedge G=G\wedge e^{-b}$ for any form G. Notice also that $e^{b}\wedge e^{-b}=1$, so that (\ref{F}) is easily inverted as
\begin{eqnarray}
dD=e^{b}\wedge F\;.   
\end{eqnarray}
}
\begin{eqnarray}
F = e^{-b}\wedge dD\;. \label{F}   
\end{eqnarray} 
As an example, the projection of (\ref{F}) to degree 5 is $F^{(5)}=dD^{(4)}-b\wedge dD^{(2)} + \frac12 b\wedge b \wedge dD^{(0)}$. This simplification is achieved after a particular choice for the potentials $D^{(n)}$ which are defined as a convenient combination of R-R potentials $C^{(n)}$ and the Kalb Ramond 2-form
\begin{eqnarray}
D^{(0)}=C^{(0)}\;,\;\;\;\;
D^{(2)}=C^{(2)} + b\wedge C^{(0)} \;,\;\;\;\; 
D^{(4)}=C^{(4)} + \frac12 b\wedge C^{(2)} + \frac12 b\wedge b\wedge C^{(0)}\;,\cr 
D^{(1)}=C^{(1)} \;, \;\;\;\;\;\;\;\;\; 
D^{(3)}= C^{(3)} +  b\wedge C^{(1)} \;, 
\;\;\;\;\;\;\;\;\;\;\;\;\;\;\;\;\;\;\;\;\;\;\;\;\;\;\;\;\;\;\;\;\;\;\;\;\;\;\;\;\;\label{Dforms}
\end{eqnarray}
while $D^{(p)}$ of higher orders are introduced as the electromagnetic duals of the previous ones, leading to the following duality relations on the corresponding curvatures
\begin{eqnarray}
* F^{(n)}= (-)^{\frac{n(n-1)}{2}} F^{(10-n)} \;,  \label{dualityR}
\end{eqnarray}
with 
\begin{eqnarray}
*F^{(n)}=\frac{1}{(10-n)!\, n!} \epsilon_{\nu_1 ... \nu_n\mu_1...\mu_{10-n}} F^{(n)}{}^{\nu_1 ... \nu_{n}} 
dx^{\mu_1}\wedge \dots \wedge dx^{\mu_{10-n}} \,,  
\end{eqnarray}
which implies (for 10D Minkowski signature) $ **F^{(n)}=(-)^{n+1} F^{(n)}$. So defined, the field strengths are subject to the following $BI$
\begin{eqnarray}
d\left(e^{b}\wedge F\right)=0 \;. \label{BI}  
\end{eqnarray}
The compatibility of both formulations works at the level of $e.o.m.$ and $BI$. It requires to vary the action by considering the full set of forms as independent fields, which leads to

\begin{eqnarray}
 \delta \left(\int d^{10}x \sqrt{-g} \; {\cal L}_{R}\right) 
 &=&
  -\frac12 \int d^{10}x \sqrt{-g}  \; \sum_{p} \frac{1}{p!}\delta D^{(p)}_{\mu_1\dots \mu_p} 
  \left(*\left[ d\left(e^{-b}\wedge *F\right)\right]\right)^{\mu_1 \dots \mu_p}  + ...\cr
 &=& -\frac12 \int\; 
 \sum_{p} (-)^{p} \delta D^{(p)}\wedge d\left(e^{-b}\wedge *F\right)  + ...
\end{eqnarray}
where ellipsis stand for total derivatives and terms proportional to $\delta g_{\mu\nu}$ or $\delta b_{\mu\nu}$. 
Hence, vanishing of the $e.o.m.$ requires
\begin{eqnarray}
d(e^{-b}\wedge * F) = 0 \;, \label{eom}
\end{eqnarray}
It is exactly at this point (after varying the action) when the degrees of freedom ($d.o.f.$) are truncated by imposing (\ref{dualityR}). 

The duality relation (\ref{dualityR}) switches $*F\to F$ in (\ref{eom}) up to a the sign $(-)^{\frac{n(n-2)}{2}}$, which can be absorbed in the exponential of the Kalb Ramond by flipping its sign. Then, $e.o.m\longleftrightarrow BI$ after (\ref{dualityR}) which reduces by half the number of independent equations, matching precisely the $e.o.m.$ and $BI$ of the standard formulation of type IIA
\begin{eqnarray}
d\left(*F^{(4)}\right)=-H\wedge F^{(4)} \;,\;\;\;\;\;\;\;\;\;\;\;\;\;\;\;\;\;\;\;\;\;\;\;\;
dF^{(4)}=-H\wedge F^{(2)} \;,\cr
d\left(*F^{(2)}\right)=H\wedge *F^{(4)} \;,\;\;\;\;\;\;\;\;\;\;\;\;\;\;\;\;\;\;\;\;\;\;
dF^{(2)}=0 \;,\;\;\;\;\;\;\;\;\;\;\;\;\;\;\;  \cr
\end{eqnarray}
and type IIB, respectively
\begin{eqnarray}
d\left(*F^{(3)}\right)=H\wedge F^{(5)} \;,\;\;\;\;\;\;\;\;\;\;\;\;\;\;\;\;\;\;\;\;\;\;\;\;
dF^{(3)}=-H\wedge F^{(1)} \;,\cr
d\left(*F^{(1)}\right)=H\wedge *F^{(3)} \;,\;\;\;\;\;\;\;\;\;\;\;\;\;\;\;\;\;\;\;\;\;\;
dF^{(1)}=0 \;,\;\;\;\;\;\;\;\;\;\;\;\;\;\;\;  \cr
d\left(* F^{(5)}\right) = dF^{(5)}=  - H\wedge F^{(3)}  \;.\;\;\;\;\;\;\;\;\;\;\;\;\;\;\;\;\;\;\;\;
\end{eqnarray}

\section{Duality action in Kaluza Klein compactifications}

In this section, we will discuss the emergence of the duality group in toroidal compactifications of the bosonic sector of type II theories, as outlined in \cite{Fukuma:1999jt}. We will proceed by performing a $GL(d)$ decomposition of the duality multiplets to uncover how the geometric and non-geometric sectors manifest in its components. This decomposition is crucial in order to establish the role of $\beta$ as an effective symmetry in the 10-dimensional framework, in next section.
\medskip 

\subsection{$O(d,d)$ and the NS-NS sector}
The T-duality group acts linearly on both NS-NS and R-R sectors after a KK reduction. This is achieved after a particular organization of the $d.o.f.$ The effective action in a toroidal compactification is obtained by taking the following decomposition of $GL(10)\to GL(n)\oplus GL(d)$, with $d=10-n$ in the NS-NS sector

\begin{eqnarray}
 \hat g_{\mu\nu} dx^{\mu} dx^{\nu} &=& g_{mn} dx^{m} dx^{n} + G_{ij}\left(dy^i + A^i{}_{m}dx^{m} \right)\left(dy^j + A^j{}_{n}dx^{n} \right)  \cr
 \frac12 \hat b_{\mu\nu} dx^{\mu}\wedge dx^{\nu} &=& 
 \frac12 \left( b_{mn} - B_{im} A^{i}{}_{n}\right) dx^{m}\wedge dx^{n} + B_{im} dx^{m}\wedge \left(dy^{i}+ A^{i}{}_{n} dx^{n}\right) \cr
 && +\frac12 B_{ij}\left(dy^{i}+ A^{i}_{m} dx^{m}\right)\wedge \left(dy^{j}+ A^{j}_{n} dx^{n}\right)\cr
 \hat\phi &=& \phi -\frac14 \log(G)\label{decomposition}
\end{eqnarray}

In this section (and only in section 3) we will use {\it hatted} notation to denote 10 dimensional tensors: $\hat g, \hat b, \hat\phi$.  
The field redefinitions in (\ref{decomposition}) ensure nice properties under the local symmetries (diff + gauge) of the effective theory in n-dimensions. On the other hand, global symmetry ($GL(d)$ and ${\mathbb R}^{d(d-1)/2}$) in the internal space is enhanced into an $O(d,d)$ group, for toroidal compactifications, due to the appearance of new symmetries induced by the isometries of the background.  Making this symmetry manifest requires to accommodate scalars and 1-forms in duality multiplets $\left({\cal M}_{MN} \right)$ ${\cal A}_{Mn}$ transforming in the (bi)fundamental representation
\begin{eqnarray}
{\cal M}=\left(\begin{matrix}
G^{ij} & -G^{ik}B_{kj}\cr
B_{ik} G^{kj} & G_{ij} - B_{ik}G^{kl} B_{lj}
\end{matrix}\right)    \;,\;\;\;\;\;\;\;\;
{\cal A} = \left(\begin{matrix}
- A^{i}{}_{m}\cr
B_{im} - B_{ik} A^{k}{}_{m}
\end{matrix}\right)  \;.
\end{eqnarray}

The subgroup connected with the identity element, $SO^{+}(d,d)$ is generated by exponentiation of the Lie algebra elements, in the fundamental representation. Its generators can be organized by those leading to a $GL(d)$ subgroup, the b-shifts and the $\beta$-transformations respectively $t_{\alpha}=\{t^{i}{}_{j}, t^{ij}, t_{ij}\}$, where the only non-vanishing components of $(t_{\alpha})_{M}{}^{N}$ are
\begin{eqnarray}
\left(t^{i}{}_{j}\right)^{k}{}_{l}= - \delta^{i}_{l}\delta^{k}_{j}\;,\;\;\;\;
\left(t^{i}{}_{j}\right)_{k}{}^{l}= \delta^{i}_{k}\delta_{j}^{l}\;;\;\;\;\;
\left(t^{ij}\right)_{kl}=\delta^{ij}_{kl}\;;\;\;\;\;
\left(t_{ij}\right)^{kl}=\delta_{ij}^{kl}\;.\;\;\;\;
\end{eqnarray}
A general element of the $SO^{+}(d,d)$ group is constructed after proper multiplication of $h_{a}, h_{b}$ and $h_{\beta}$, where
\begin{eqnarray}
h_{a}=e^{a_{i}{}^{j}t^{i}{}_{j}}=\left(\begin{matrix}
e^{-a^T} & 0 \cr 0 & e^{a}     
\end{matrix}\right)     \; ,\;\;\;\;\;\;\;
h_{\mathfrak{b}}=e^{\mathfrak{b}_{ij}t^{ij}}=\left(\begin{matrix}
1 & 0 \cr \mathfrak{b} & 1     
\end{matrix}\right)     \; ,\;\;\;\;\;\;\;
h_{\beta}=e^{\beta^{ij} t_{ij}}= \left(\begin{matrix}
1 & \beta \cr 0 & 1     
\end{matrix}\right)     \; ,\;\;\;\;\;\;\;\label{G-b-beta}
\end{eqnarray}
One readily confirms that the transformation ${\cal M}\to {\cal M}'=h {\cal M}h^T$ leads to
\begin{eqnarray}
\delta_a G_{ij} &=& 2 a_{(i}{}^{k} G_{|k| j)} \;,\;\;\;\;\;  
\delta_{\mathfrak{b}} G_{ij} = 0 \;,\;\;\;\;\; 
\delta_{\beta} G_{ij} = - 2 B_{(i|k|} \beta^{k}{}_{j)} \;,\;\cr
\delta_{a} B_{ij} &=& 2 a_{[i}{}^{k} B_{|k| j]} \;,\;\;\;\;\; 
\delta_{\mathfrak{b}} B_{ij} = \mathfrak{b} \;,\;\;\;\;\; 
\delta_{\beta} B_{ij} = - 2 \beta_{ij} - B_{ik} \beta^{kl} B_{lj}\;.\;\;\;\;\; \label{betaScalars}
\end{eqnarray}
Similarly, the transformation of ${\cal A}$ reads, in components
\begin{eqnarray}
\delta_a A^{i}{}_{m} &=& -a_{j}{}^{i} A^{j}{}_{m} \;,\;\;\;\;     
\delta_{\mathfrak{b}} A^{i}{}_{m} = 0 \;,\;\;\;\;     
\delta_{\beta} A^{i}{}_{m} =  - \beta^{ij} \left(B_{jm} - B_{ik} A^{k}{}_{m}\right)\;,\;\;\;\;     \cr
\delta_a B_{i}{}_{m} &=& a_{i}{}^{j} B_{j}{}_{m} \;,\;\;\;\;     
\delta_{\mathfrak{b}} B_{i}{}_{m} = 0\;,\;\;\;\;     
\delta_{\beta} B_{i}{}_{m} = - \beta_{ik} A^{k}{}_{m} - B_{ik} \beta^{kj} B_{jm}  \;,\;\;\;\;\label{betaVectors}     
\end{eqnarray}
where indices in $\beta$ are lowered with the internal metric $G_{kl}$. It is worth noticing that while the $GL(d)$ and b-shift of the compact space are trivially promoted to the whole 10-dimensional space, the main reason for $\beta$ to admit a promotion to higher dimensions upon (\ref{betaConstraint}), is because (\ref{betaScalars}) and (\ref{betaVectors}) plugged with (\ref{decomposition}) implies the following embedding in 10d, 
\begin{eqnarray}
\delta_\beta \hat{g}_{\mu\nu}= 
- \hat{b}_{\mu \rho} \hat{\beta}^{\rho\sigma} \hat{g}_{\sigma \nu}
- \hat{b}_{\nu \rho} \hat{\beta}^{\rho\sigma} \hat{g}_{\sigma \mu}\;,\;\;\;\;\;     
\delta_\beta \hat{b}_{\mu\nu}= - \hat{g}_{\mu \rho} \hat{\beta}^{\rho\sigma} \hat{g}_{\sigma \nu} - \hat{b}_{\mu \rho} \hat{\beta}^{\rho\sigma} \hat{b}_{\sigma \nu}  \;,\;\;\;\;\;  
\end{eqnarray}
where $\hat{\beta}^{ij}=\beta^{ij}, \hat{\beta}^{mi}=\hat{\beta}^{im}= \hat{\beta}^{mn}=0$ and this solution for $\hat \beta$ is precisely a consequence of the condition (\ref{betaConstraint}).

\subsection{$O(d,d)$ and the R-R sector}

Let us move now to the R-R sector. The global symmetry in this sector admits a linear realization because p-forms of the democratic formulation naturally fit in a Majorana-Weyl representation of $SO(d,d)$ using fermionic operators $\psi_{i}$. 

 Let us very briefly comment on this construction (for further details see \cite{Fukuma:1999jt}). Let $\psi_{i}$, with $\psi^{i}{}^{\dagger}:=(\psi_i)^{\dagger}$, be a set of d dimensional fermionic operators satisfying an anti-commuting algebra
 \begin{eqnarray}
  \{\psi_{i},\psi^{j}{}^{\dagger}\}=\delta_{i}{}^{j}\; {\mathbb I}  \; ,\;\;\;\;\;  \{\psi_{i},\psi_{j}\}=0 \;,  \;\;\;\;\; \{\psi^{i}{}^{\dagger},\psi^{j}{}^{\dagger}\}=0  \;.
 \end{eqnarray}
 An $O(d,d)$ Clifford algebra is built after identifying $\Gamma_{i}=\sqrt{2} \psi_{i}, \Gamma^{i}=\sqrt{2} \psi^{i}{}^{\dagger}$
 \begin{eqnarray}
 \{\Gamma_{M},\Gamma_{N}\}= 2\eta_{MN}\; {\mathbb I}\;,
 \;\;\;\;\;\;\;\;\;\;\;
  \eta_{MN}=\left(\begin{matrix}
 0 & \delta^{i}{}_{j}\cr \delta_{i}{}^{j} & 0    
 \end{matrix}\right)  \;,  
 \end{eqnarray}
 where $\eta$ is the $O(d,d)$ invariant metric. Hence, there is a natural isomorphism between the fermionic Fock space and p-forms of the internal space (duality group does not act on the non-compact directions and so (10-d) indices are implicit in $D$)\footnote{Here lower and upper index $(p)$ characterizes the degree of the form in the compact and non-compact spaces, respectively.}
 \begin{eqnarray}
\ket{D} =\sum_{p} \frac{1}{p!} D_{(p)i_1\dots i_p} \psi^{i_1}{}^{\dagger} \dots \psi^{i_p}{}^{\dagger} \ket{0} \;,   \label{Cspinor}
 \end{eqnarray}
$\ket{0}$ being the vacuum of the Fock space, satisfying 
\begin{eqnarray}
\psi_{i}\ket{0}=0\;,\;\;\;\;\;    \bra{0}\ket{0}=1\;.
\end{eqnarray}
 
  The sequence of field redefinitions in the R-R sector is as follows. First, each p-form $f^{(p)}$ splits into a sum of q-forms $f^{(q)}_{(n)i_1\dots i_n}$
 \begin{eqnarray}
     f^{(p)}= \sum_q \frac{1}{n!} f^{(q)}_{(n)i_1 \dots i_n} dy^{1}\wedge \dots \wedge dy^{n}\;,
 \end{eqnarray}
where $n=p-q$ and each $f^{(q)}_{(n)i_1 \dots i_n}$ is a q-form in the non-compact space
\begin{eqnarray}
    f^{(q)}_{(n)i_1 \dots i_n}= \frac{1}{q!}
    f^{(q)}_{m_1\dots m_q i_1 \dots i_n}
    dx^{m_1}\wedge\dots \wedge dx^{m_q}\;.
\end{eqnarray}
Then, one introduces shifted curvatures $F'$, formally defined as
\begin{eqnarray}
F'=\left.F\vphantom{2^{\frac12}}\right|_{dy^i\to dy^i-A^i} \;.   
\end{eqnarray}
This redefinition simplifies the kinetic terms $\sum_p |F^{(p)}|^2$ after the KK decomposition by absorbing the dependence on the metric 1-forms $A^{i}=A^{i}{}_{m} dx^m$.  

The action of the duality subgroup connected to the identity, the $Spin^{+}(d,d)$ group,\footnote{$O(d,d)$ is the maximum subgroup of the U-duality group that closes independently on the NS-NS and R-R sectors. Nevertheless, the T-duality transformations around single circles (represented by matrices with negative determinant) do not preserve chirality, so only the $SO(d,d)$ lift $Spin(d,d)$ is separately a symmetry of type IIA and type IIB, after compactification.} is generated with the spinorial representation of elements in (\ref{G-b-beta}), 
\begin{eqnarray}
S_a=\frac{1}{\sqrt{det(A)}}e^{a_{i}{}^{j}\psi^i{}^{\dagger}\psi_{j} }   \;,\;\;\;\;\;
S_{\mathfrak{b}}=e^{\frac12 \mathfrak{b}_{ij}\psi^i{}^{\dagger} \psi^j{}^{\dagger} }   \;,\;\;\;\;\;
S_{\beta}=e^{-\frac12 \beta^{ij}\psi_i\psi_{j} } \;,\;\;\;\;\;
\end{eqnarray}
where $A_{i}{}^{j}=(e^{a})_i{}^j$.

Therefore, after (\ref{Cspinor}) we find the following transformation rules for the internal p-forms $D_{(p)i_1\dots i_p}$ 
\begin{eqnarray}
\delta_{a}D_{(p)i_1 \dots i_p} &=& p a_{[i_1}{}^{k}\; D_{(p)|k|i_2 \dots i_p]} 
-\frac12 a_{k}{}^{k} D_{(p) i_1 \dots i_p}\;, \cr
\delta_{\mathfrak{b}} D_{(p)i_1 \dots i_p} &=&  \left(\mathfrak{b}\wedge D_{(p-2)}\right)_{i_1 \dots i_p} =  \frac{p!}{2!(p-2)!} \mathfrak{b}_{[i_1 i_2}D_{(p-2)i_3 \dots i_p]} \;,\cr 
\delta_\beta D_{(p)i_1 \dots i_p} &=& -\frac12 \left(i_{\beta} D_{(p+2)} \right)_{i_1 \dots i_p} = -\frac12 \beta^{kl} D_{(p+2)kli_1 \dots i_p} \;,
\end{eqnarray}
where in the last line we have introduced the interior product of $\beta$ with (p+2)-forms.

\section{$\beta$ symmetry of the RR sector in 10d}

After analyzing how an infinitesimal duality transformation acts on the internal p-forms after a Kaluza-Klein reduction, we proceed similarly to the NS-NS sector. We promote these transformation rules as an effective symmetry of the parent theory, subject to the constraint (\ref{betaConstraint}) 
\begin{eqnarray}
\delta_{a}D^{(p)}_{\mu_1 \dots \mu_p} &=& p a_{[\mu_1}{}^{\nu}\; D^{(p)}_{|\nu|\mu_2 \dots \mu_p]}\;- \frac12 a_{\mu}{}^{\mu} D^{(p)}_{\mu_1 \dots \mu_p}; ,  \label{daD} \\
\delta_{\mathfrak{b}} D^{(p)}_{\mu_1 \dots \mu_p} &=&  \left(\mathfrak{b}\wedge D^{(p-2)}\right)_{\mu_1 \dots \mu_p} \;,\label{dbD}\\ 
\delta_\beta D^{(p)}_{\mu_1 \dots \mu_p} &=& -\frac12 \left(i_{\beta} D^{ (p+2)}\right)_{\mu_1 \dots \mu_p} \;. \label{dbetaD} 
\end{eqnarray}
In this expression $D^{(p)}$ is assumed to be non vanishing only for $p=0,...,8$. Global $GL(10)$ acts on RR p-forms with a non trivial density contribution (last factor in (\ref{daD})). While the linear piece (first term in the $r.h.s$ of (\ref{daD})) cancels the variation of the inverse metric in any coupling with indices properly contracted, the density piece cancels the contribution of the measure $\delta_a({\sqrt{-g}})$ as long as the couplings are quadratic in $D^{(p)}$.\footnote{Interestingly, the field redefinition $\tilde D^{(p)}= e^{\phi} D^{(p)}$ transforms linearly (without density factor) under $GL(10)$ and the action written with these variables has the usual (duality singlet) measure $\sqrt{-g}e^{-2\phi}$. This new scheme makes explicit the dilaton dependence of the string loop expansion, at the cost of introducing dilaton dependence on the gauge transformations and Bianchi identities.} b-shifts are also a manifest symmetry in 10 dimensions if the action is written in terms of the curvatures (\ref{F}). Indeed
\begin{eqnarray}
    \delta_{\mathfrak{b}} F &=& \delta_{\mathfrak{b}}\left(e^{-b}\wedge dD\right)= -\delta_{\mathfrak{b}}b \wedge e^{-b}\wedge dD + e^{-b}\wedge d(\delta_{\mathfrak{b}}D)\cr
    &=&-\mathfrak{b} \wedge e^{-b}\wedge dD + e^{-b}\wedge d(\mathfrak{b}\wedge D) =0.
\end{eqnarray}

We focus now on the non geometric sector of $O(d,d)$ described by $\beta$. We remark once again that while (\ref{daD}) and (\ref{dbD}) are symmetries of the parent theory, we expect (\ref{dbetaD}) to be a symmetry only in an effective way, when it is supplemented with the constraint (\ref{betaConstraint}).  

An important observation is that the $\beta$-transformations of the NS-NS sector in the KK reduction remain unchanged even when RR-potentials are introduced. Based on this, we propose that equation (\ref{NSbeta}) is not affected or deformed. Consequently, equation (\ref{dLNS}) still holds, indicating that both ${\cal L}_{NS}$ and ${\cal L}_{R}$ should remain invariant independently.

In this section, we will provide a detailed verification that the $\beta$ transformations, as defined in equation (\ref{dbetaD}), are indeed a symmetry of (\ref{LR}). Based on the previous discussion, it is evident that any combination of $\sqrt{-g}\;|F^{(p)}|^2$ is separately invariant under $GL(10)$ and b-shifts in 10 dimensions. We will show that, in analogy with the NS-NS sector, $\beta$ transformations constrain the relative coefficients of the kinetic terms. To accomplish this,  we start by computing the induced transformation on the field strengths $F^{(p)}$, due to (\ref{dbetaD}),
\begin{eqnarray}
\delta_\beta F 
&=&- \delta_\beta b \wedge e^{-b} \wedge dD 
-\frac12 e^{-b} \wedge [i_{\beta}(e^b\wedge F)] \;, \label{deltaF}
\end{eqnarray}
where we have used
\begin{eqnarray}
\delta(dD)&=&\sum_p \delta \left( dD^{(p)}\right) = \sum_{p} \frac{1}{p!}\delta_\beta\left(\partial_{[\mu_1}D^{(p)}_{\mu_2 \dots \mu_{p+1}]} \right) dx^{\mu_1}\wedge \dots \wedge dx^{\mu_{p+1}} \cr
&=& \sum_{p} 
-\frac12 \frac{1}{p!} \partial_{[\mu_1} \left(\beta^{\rho\sigma} D^{(p+2)}_{\rho\sigma\mu_2 \dots \mu_{p+1}]} \right) dx^{\mu_1}\wedge \dots \wedge dx^{\mu_{p+1}}\cr
&=& \sum_{p} -\frac12 \frac{1}{p!}\beta^{\rho\sigma} \frac{(p+3)}{(p+1)}{}\partial_{[\rho}D^{(p+2)}_{\sigma\mu_1 \dots \mu_{p+1}]} dx^{\mu_1}\wedge \dots \wedge dx^{\mu_{p+1}}\cr
&=& \sum_{p} -\frac12 i_{\beta}\left(dD^{(p+2)}\right) = -\frac{1}{2} i_{\beta}(dD) \;,
\label{deltadD}
\end{eqnarray}
where condition (\ref{betaConstraint}) was used in the third line. Hence, (\ref{deltaF}) follows from (\ref{deltadD}), after inverting equation (\ref{F}). 

Expression (\ref{deltaF}) can be further manipulated by noting that 
\begin{equation}
    i_{\beta}[A^{(p)}\wedge B^{(q)}] = i_\beta (A^{(p)})\wedge B^{(q)} + A^{(p)}\wedge i_{\beta}(B^{(q)}) + 2[A\beta B]\;,\label{ibetaAB}
\end{equation}
where $A^{(p)}$ and $B^{(q)}$ are two arbitrary forms. Here we have defined
\begin{eqnarray}
    [A\beta B] = \frac{1}{(p-1)!}\frac{1}{(q-1)!}A^{(p)}_{[i_1\dots i_{p-1}|k}\beta^{kl}B^{(q)}_{l|i_p\dots i_{p+q-2}]} dx^{i_1}\wedge \dots\wedge dx^{i_{p+q-2}}\;.\label{AbetaB}
\end{eqnarray}
Interestingly, we can use (\ref{ibetaAB}) inductively in order to obtain 
\begin{eqnarray}
i_\beta (e^{b}) &=& i_\beta\left(\sum_{m\geq0} \frac{1}{m!}b\wedge \dots \wedge b\right) \cr &=& e^{b} \wedge \left(\vphantom{\frac12} i_\beta b + 2[b\beta b]\right)\;.\label{ibetaeb}
\end{eqnarray}
Similarly we can use (\ref{AbetaB}) inductively to get 
\begin{eqnarray}
[e^{b}\beta F]= e^{b}\wedge [b\beta F].    
\end{eqnarray}
So, the second term in (\ref{deltaF}) can therefore be expressed as 
\begin{eqnarray}
-\frac12 e^{-b} \wedge [i_{\beta}(e^b\wedge F)] &=&
-\frac12 e^{-b} \wedge \left( i_\beta (e^{b})\wedge F + e^{b}\wedge i_{\beta}F + 2[e^{b}\beta F] \right)\cr
&=& -\frac12 \left(\vphantom{\frac12} (i_\beta b )\,  F +2 [b\beta b]\wedge F + i_{\beta}F + 2[b\beta F] \right) \;, \label{deltaF2}
\end{eqnarray}
where we have used $e^{-b}\wedge e^{b} =1$. The first term in (\ref{deltaF}) can be written instead as
\begin{eqnarray}
    - \delta_\beta b \wedge e^{-b} \wedge dD = 
    \left(\beta + [b\beta b]\right) \wedge F\;.\label{deltaF1}
\end{eqnarray}
Here we have introduced 
\begin{eqnarray}
\beta = \frac12 \beta_{\mu\nu} dx^{\mu}\wedge dx^{\nu} \;,
\end{eqnarray}
with $\beta_{\mu\nu}=\beta^{\rho\sigma}g_{\mu\rho} g_{\nu\sigma}$. Plugging (\ref{deltaF2}) and (\ref{deltaF1}) we finally obtain
\begin{eqnarray}
\delta F =  -\frac12  (i_\beta b )\,  F - [b\beta F] + \beta  \wedge F-\frac12  i_{\beta}F \;,
\end{eqnarray}
or in components
\begin{eqnarray}
\delta F^{(p)}_{\mu_1 \dots \mu_p} &=&  -\frac12  \beta^{\rho\sigma} b_{\rho\sigma}   F^{(p)}_{\mu_1 \dots \mu_p} - p \;b_{[\mu_1|\rho|}\beta^{\rho\sigma}F^{(p)}_{|\sigma|\mu_2 \dots \mu_p]} \cr
&+& \frac{p (p-1)}{2} \beta{\vphantom{F^{(p)}_\mu}}_{[\mu_1 \mu_2}  F^{(p-2)}_{\mu_3 \dots \mu_p]} -\frac12  \beta^{\rho\sigma} F^{(p+2)}_{\rho\sigma\mu_1 \dots \mu_p} \;. \label{deltaFcomponents}
\end{eqnarray}
In this expression, it is implicitly assumed that $F^{(p)}$ is non vanishing only for $p=1,3,5,7,9$ in type IIB and $p=2,4,6,8$ for type IIA. 

With these results we are ready to analyze the invariance of the Lagrangian ${\cal L}_{R}$.

\subsection{$\beta$-invariance of the Lagrangian in type IIA and type IIB} \label{sec4.1}

The variation of the R-R sector of the Lagrangian due to a $\beta$ transformation is
\begin{eqnarray}
\delta_{\beta}\left[\sqrt{-g}{\cal L}_{R}\right]   &=&
\sqrt{-g}\left[\vphantom{{\frac12}^{\frac12}}\right.\frac{\delta_{\beta}\left(\sqrt{-g}\right)}{\sqrt{-g}} {\cal L}_R + \sum_{p} \frac{p}{p!} \delta_{\beta}g^{\mu_1\nu_1} F^{(p)}_{\mu_1 \mu_2 \dots \mu_p} F^{(p)}_{\nu_1}{}^{\mu_2 \dots \mu_p}
\cr
&& \;\;\;\;\;\;\;
+ \sum_{p} \frac{2}{p!} F^{(p)}{}^{\mu_1 \dots \mu_p} \delta_{\beta}F^{(p)}_{\mu_1 \dots \mu_p}
\left.\vphantom{{\frac12}^{\frac12}}\right]\;,\label{deltaLR}
\end{eqnarray}
with the sum runing over $p=1,3,5,7,9$ ($p=2,4,6,8$) on type IIB (IIA). Replacing the transformation of the volume measure and inverse metric in (\ref{deltaLR}):
\begin{eqnarray}\label{eq:deltaMetrica}
\delta \sqrt{-g} &=& \sqrt{-g} \; i_{\beta} b \;, \cr
\delta g^{\mu\nu} &=& 2 \beta^{\rho(\mu} b^{\nu)}{}_{\rho}\;,
\end{eqnarray}
one readily verifies that first line in (\ref{deltaLR}) exactly cancels terms in $\delta F^{(p)}$ proportional to $F^{(p)}$ (first line in (\ref{deltaFcomponents})), leaving only mixed terms
\begin{eqnarray}\label{eq:Cancelation}
\delta_{\beta}\left[\sqrt{-g}{\cal L}_{R}\right] \;  = \hspace{12cm} \cr
\sqrt{-g}\sum_{p} \left(\frac{p (p-1)}{2\; p!} \beta{\vphantom{F^{(p)}_\mu}}_{\mu_1 \mu_2}  F^{(p-2)}_{\mu_3 \dots \mu_p}F^{(p)}{}^{\mu_1 \dots \mu_p} -\frac{1}{2\; p!}  \beta^{\rho\sigma} F^{(p+2)}_{\rho\sigma\mu_1 \dots \mu_p} F^{(p)}{}^{\mu_1 \dots \mu_p} \right)
= 0\;,\;\;\;\;\;
\end{eqnarray}
where we immediately note that terms cancel each other out in the sum over all form orders. As we anticipated, it is now clear from this analysis that invariance of ${\cal L}_{R}$ is broken if we modify the relative coefficients of the kinetic terms.

\subsection{$\beta$-invariance of the Lagrangian in massive type IIA} \label{sec4.2}

Massive type IIA is a deformation of type IIA supergravity containing extra curvatures $F^{(0)}$ and $F^{(10)}$. The whole set of field strengths are defined like in (\ref{F}) by adding formally a 0-form to $dD$
\begin{eqnarray}
F=e^{b}\wedge(dD + M) \;.    \label{FM}
\end{eqnarray}
So, $F^{(0)}=M$ and the remaining forms contain an extra contribution $e^{b} M$. 

The Lagrangian of the RR sector is still (\ref{LR}) with the sum now running over all allowed even degrees $n=0,2,\dots,10$. We remark here that while $F^{(10)}$ can be interpreted as the curvature of a $D^{(9)}$ form potential, the field  $F^{(0)}$ does not have such an interpretation and is formally introduced in (\ref{LR}) and finally interpreted as the hodge dual of the former.\footnote{A formal (-1)-form potential for $F^{(0)}$ was introduced in \cite{Lavrinenko:1999xi}. Its mathematical interpretation was further clarified in \cite{Hohm:2011cp}, as a contribution of the 1-form potential having a linear dependence on a coordinate dual to a winding mode.} Therefore, the new fields carry non propagating degrees of freedom. Certainly, the $e.o.m.$ of $D^{(9)}$ is
\begin{eqnarray}
d(*F^{(10)})=0\;,\label{eomD9}
\end{eqnarray}
but being $*F^{(10)}$ a 0-form, equation (\ref{eomD9}) implies $*F^{(10)}$ ($=F^{(0)}$ after (\ref{dualityR})) =constant. Despite this observation, it is worth noting that massive type II supergravity does indeed arise in type IIA string. The 9-form potential is not observed in the quantization of the type IIA string precisely because it does not carry propagating $d.o.f.$, but it is required by the very existence of D8-branes. 

We now turn to explore whether $\beta$-symmetry can be extended in this case as well. We immediately see that $\delta D^{(9)}$ is trivially invariant if (\ref{dbetaD}) holds, because it is defined through the contraction with an hypothetical 11-dimensional form in 10d, but there is no interpretation at the level of potential forms for $F^{(0)}$. For that reason we work directly at the level of curvatures and extrapolate (\ref{deltaFcomponents}), instead of (\ref{dbetaD}). 

With this definition it readily follows that $\beta$ symmetry is consistent with constant $F^{(0)}=M$, as (\ref{deltaFcomponents}) implies
\begin{eqnarray}
\delta_{\beta} F^{(0)} = -\frac12 \beta^{\rho\sigma} b_{\rho\sigma} F^{(0)} -\frac12 \beta^{\rho\sigma} F^{(2)}_{\rho\sigma}  = -\frac12 \beta^{\rho\sigma} \partial_{\rho}D^{(1)}_{\sigma} =0 \;.\label{Minvariant}
\end{eqnarray}

The M-dependent deformation on the curvatures $F^{(p)}$, for $p=2,4,6,8$ (\ref{FM}) and the introduction of curvatures $F^{(0)}$ and $F^{(10)}$ in (\ref{LR}) do not spoil the proof of invariance of the action of the previous section, which only depends on the formal expression (\ref{deltaFcomponents}). 

We would like to emphasize here that massive type IIA can be considered as a consistent truncation of type IIA with respect to infinitesimal $O(d,d)$ transformations. This is straightforwardly verified for $GL(d)$ and {\it b-shift}. Regarding $\beta$ transformation we see that $M=0$ is also consistent due to (\ref{Minvariant}), while $F^{(10)}$ transforms non trivially 
\begin{eqnarray}
\delta_{\beta} F^{(10)}_{\mu_1 \dots \mu_{10}} = -\frac12 \beta^{\rho\sigma} b_{\rho\sigma} F^{(10)}_{\mu_1 \dots \mu_{10}} - 10 b_{[\mu_1|\rho|}\beta^{\rho\sigma} F^{(10)}_{|\sigma|\mu_2 \dots \mu_{10}]} + 45 \beta_{[\mu_1\mu_2} F^{(8)}_{\mu_3\dots \mu_{10}]}  \label{deltaF10}    
\end{eqnarray}
and so it could rise doubts about the consistency of the truncation as the condition $F^{(10)}=0$, $\delta_\beta F^{(10)}=0$ requires vanishing of the last factor in (\ref{deltaF10}), or equivalently
\begin{eqnarray}
\beta^{[\mu_1\mu_2} F^{(8)}{}^{\mu_3\dots \mu_{10}]}=0 \; , \label{betaF8}  
\end{eqnarray}
a condition which is not satisfied, in general. Actually, there is no conflict here as  $D^{(9)}=0$ does not imply $F^{(10)}=0$. We can still introduce a non vanishing $F^{(10)}$ in the democratic formulation of (massless) type IIA SUGRA through (\ref{F}), which transforms non trivially. Of course, this curvature is therefore interpreted as a particular combination of Kalb-Ramond and curvatures $F^{(p)}$, for $p=2,4,6,8$ and so it carries no independent $d.o.f.$ The condition $F^{(10)}=0$ is only a consequence of $F^{(0)}=M=0$, after and only after the relation $F^{(10)}=*F^{(0)}$ is imposed. Indeed, when duality relations are required to hold, (\ref{betaF8}) turns into 
\begin{eqnarray}
0=\beta^{[\mu_1\mu_2} \epsilon^{\mu_3\dots \mu_{10}]\rho\sigma} F^{(2)}_{\rho\sigma}= 
\beta^{[\mu_1\mu_2} \epsilon^{\mu_3\dots \mu_{10}]\rho\sigma} \left(\partial_{\rho}D^{(1)}_{\sigma} - M b_{\rho\sigma}\right)\;\;.  
\end{eqnarray}
Last factor cancels for $M=0$, while the first one cancels due to (\ref{betaConstraint}), because the index $\rho$ in the derivative necessarily matches one of the indices in $\beta$.

\subsection{Consistency with duality relations}\label{sec4.3}
In this section we will analyze compatibility of beta transformations and duality relations. Concretely, we want to verify that both sides in (\ref{dualityR}) transform consistently. The dual curvatures transform as
\begin{eqnarray}
   \delta [* F^{(q)}]_{\mu_1 \dots \mu_{10-q}} &=& \delta\left[ \frac{1}{q!} \epsilon_{\nu_1\dots \nu_q \mu_1 \dots \mu_{10-q}}   F^{(q)}{}^{\nu_1\dots \nu_q}\right] \cr
   &=& \frac{1}{q!}\epsilon_{\nu_1\dots \nu_q \mu_1 \dots \mu_{10-q}} \left[\vphantom{\frac{e^2}{2}}\right. \frac{\delta \sqrt{-g}}{\sqrt{-g}}  g^{\nu_1 \rho_1}\dots g^{\nu_q \rho_q} F^{(q)}_{\rho_1\dots \rho_q}
   \cr
   && + \; q  \left(\delta g^{\nu_1 \rho_1}\right)\dots g^{\nu_q \rho_q} F^{(q)}_{\rho_1\dots \rho_q}
    + \;  g^{\nu_1 \rho_1}\dots g^{\nu_q \rho_q} \delta F^{(q)}_{\rho_1\dots \rho_q}
   \left.\vphantom{\frac{e^2}{2}}\right] \;.\label{delta*F0}
\end{eqnarray}
The contribution in the second line in (\ref{delta*F0}) originates from the variation of the Levi-Civita tensor and it leads to 
\begin{eqnarray}
\delta [* F^{(q)}]_{\mu_1 \dots \mu_{10-q}}\supset (i_{\beta} b) [*F^{(q)}]_{\mu_1 \dots \mu_{10-q}}  \;.\label{Aux1delta*F}  
\end{eqnarray}
First factor in the third line in (\ref{delta*F0}) is instead
\begin{eqnarray}
\delta [* F^{(q)}]_{\mu_1 \dots \mu_{10-q}}\supset  \frac{q}{q!} \epsilon_{\nu_1\dots \nu_q \mu_1 \dots \mu_{10-q}} \left( \beta^{\tau \nu_1} b_{\sigma \tau} + \beta^{\tau}{}_{\sigma} b^{\nu_1}{}_{\tau}\right) F^{(q)}{}^{\sigma \nu_2 \dots \nu_q} \; ,\label{Aux2delta*F}
\end{eqnarray} 
while last contribution in (\ref{delta*F0}) is, after use of (\ref{deltaFcomponents})
\begin{eqnarray}
\delta [* F^{(q)}]_{\mu_1 \dots \mu_{10-q}}&\supset& 
-\frac12 (i_{\beta} b) [*F^{(q)}]_{\mu_1 \dots \mu_{10-q}} 
+\frac12 [i_{\beta}(*F^{(q-2)})]_{\mu_1 \dots \mu_{10-q}} 
\cr
&& -\;  
\frac{q}{q!}\epsilon_{\nu_1\dots \nu_q \mu_1 \dots \mu_{10-q}} \;
\;\beta^{\tau}{}_{\sigma} b^{\nu_1}{}_{\tau}F^{(q)}{}^{\sigma\nu_2 \dots \nu_q} \cr
&& -\;  
\frac12 \frac{1}{q!}\epsilon^{\nu_1\dots \nu_q}{}_{\mu_1 \dots \mu_{10-q}} \;   \beta^{\rho\sigma} F^{(q+2)}_{\rho\sigma\nu_1 \dots \nu_p} \;.\label{Aux3delta*F}
\end{eqnarray} 

Plugging (\ref{Aux1delta*F})-(\ref{Aux3delta*F}) with (\ref{delta*F0}) we get

\begin{eqnarray}
    \delta [* F^{(q)}]_{\mu_1 \dots \mu_{10-q}}&=& 
\frac12 \beta^{\rho\sigma} b_{\rho\sigma} [*F^{(q)}]_{\mu_1 \dots \mu_{10-q}} 
+\frac12 \beta^{\rho\sigma}[*F^{(q-2)}]_{\rho\sigma \mu_1 \dots \mu_{10-q}} 
\cr
&& 
+\frac{q}{q!} \epsilon^{\nu_1\dots \nu_q}{}_{\mu_1 \dots \mu_{10-q}} \beta^{\tau}{}_{ \nu_1} b^{\sigma}{}_{\tau} F^{(q)}{}_{\sigma \nu_2 \dots \nu_q}
\cr
&& -\;  
\frac12 \frac{1}{q!}\epsilon^{\nu_1\dots \nu_q}{}_{\mu_1 \dots \mu_{10-q}} \;   \beta^{\rho\sigma} F^{(q+2)}_{\rho\sigma\nu_1 \dots \nu_q}
\end{eqnarray}
or equivalently
\begin{eqnarray}
    \delta [* F^{(q)}]_{\mu_1 \dots \mu_{10-q}}&=& 
-\frac12 \beta^{\rho\sigma} b_{\rho\sigma} [*F^{(q)}]_{\mu_1 \dots \mu_{10-q}} 
+\frac12 \beta^{\rho\sigma}[*F^{(q-2)}]_{\rho\sigma \mu_1 \dots \mu_{10-q}} 
\cr
&& 
- (10-q)  b_{[\mu_1 |\rho|} \beta^{\rho \sigma}  [*F^{(q)}]_{|\sigma|\mu_2 \dots \mu_{10-q}]} 
\cr
&& -\;  
\frac{(10-q)(9-q)}{2} \beta_{[\mu_1\mu_2} [*F^{(q+2)}]_{\mu_3 \dots \mu_{10-q}]}\;,  \label{delta*F}
\end{eqnarray}
where we have used
\begin{eqnarray}
F^{(n)}_{\rho_1 \dots \rho_{n}} &=& (-)^{n+1} [**F^{(n)}]_{\rho_1 \dots \rho_{n}}  \cr
&=& (-)^{n+1} \frac{1}{(10-n)!}\epsilon_{\sigma_{1}\dots \sigma_{10-n} \rho_1 \dots\rho_{n}}[*F^{(n)}]^{\sigma_1 \dots \sigma_{10-n}}   
\end{eqnarray}
and 
\begin{eqnarray}
\epsilon^{\rho_1 \dots \rho_{10-n} \mu_{1} \dots \mu_{n}} 
\epsilon_{\rho_1 \dots \rho_{10-n} \nu_{1} \dots \nu_{n}} =  - n! (10-n)! \delta^{[\mu_1}_{\nu_1}\dots \delta^{\mu_{n}]}_{\nu_{n}} \;,  
\end{eqnarray}
to write the right hand side in (\ref{delta*F}) entirely in terms of dual forms. It is now straightforward to verify that $\beta$ symmetry of curvatures (\ref{deltaFcomponents}) and dual curvatures (\ref{delta*F}) are consistent with the duality relation (\ref{dualityR}). Indeed, using it on both sides of (\ref{delta*F}) and identifying $p=10-q$ one easily recovers (\ref{deltaFcomponents}).    

\subsection{Equations of motion}\label{sec4.4}

We will now focus on the behavior of the $e.o.m.$ under the non geometric sector of the duality group. The variation of the action is, after integrating by parts 
\begin{eqnarray}
\delta S= \int d^{10}x \left(-2 \sqrt{-g} {\cal L}_{NS}\; \delta d + \tilde {\cal E}^{ab} \; \delta e_{ab} 
+ \tilde {\cal B}^{ab}\;  \delta b_{ab}  +\sqrt[4]{-g}\; \sum_{p} \tilde{\cal E}^{(p)}{}^{a_1 \dots a_p} \; \delta D^{(p)}_{a_1 \dots a_p}\right)\;.\label{deltaS}
\end{eqnarray}
Here we have considered the generalized dilaton $d=\phi -\frac12 \log\sqrt{-g}$ as independent $d.o.f.$\footnote{This is a convenient choice when dealing with T-duality because the generalized dilaton $d$ is a singlet. For instance, first term in (\ref{deltaS}) does not contribute to $\delta_\beta S$.} and we have introduced $\delta e_{ab} =e^{\mu}{}_{a}\delta e_{\mu b}$, $\delta b_{ab}= e^{\mu}{}_{a} e^{\nu}{}_{b} \delta b_{\mu\nu}$, $\delta D^{(p)}_{a_1 \dots a_p}= e^{\mu_1}{}_{a_1} \dots e^{\mu_p}{}_{a_p} \delta D^{(p)}_{\mu_1 \dots \mu_p}$ and
\begin{eqnarray} \label{eq:Motion}
\tilde{\cal E}_{ab} &=& - 2 \sqrt{-g} e^{-2\phi}\left(\vphantom{\frac12}  R_{a b} + 2 \nabla_{a}\nabla_{b}\phi -\frac14 H_{a c d} H_{b}{}^{c d} \right) \cr
&& \;\;\;\;\;\;\;\;\;\; 
-\frac14  \sqrt{-g} \sum_{p} \left(g_{ab} |F^{(p)}|^2 - \frac{2}{(p-1)!} F^{(p)}_{a c_1 \dots c_{p-1}} F^{(p)}_{b}{}^{ c_1 \dots c_{p-1}} \right)\, ,  
\nonumber \\
\tilde{\cal B}_{a b} &=& \sqrt{-g} e^{-2\phi}\left(\frac12 \nabla_{c} H_{ab}{}^c{}- \nabla_{c}\phi H_{ab}{}^{c} \right)
 +\frac14 \sqrt{-g} \sum_{p} \frac{1}{(p-2)!}F^{(p)}_{a b c_1 \dots c_{p-2}} F^{(p-2)}{}^{c_1 \dots c_{p-2}}\, \;,  \nonumber\\
 \tilde{\cal E}^{(p)}_{a_1 \dots a_p} &=&  -\frac12 \frac{1}{p!} \sqrt[4]{-g}\; \left(*d\left[e^{-b}\wedge *F\right]\right)_{a_1 \dots a_p} \;, 
\end{eqnarray}
$\nabla_{c}$ denotes the {\it flattened} (torsion free) covariant derivative, and the unusual power in the measure in $\tilde{\cal{E}}^{(p)}$ is chosen for later convenience. Interestingly, the non geometric sector of $O(d,d)$ transforms the vielbein equations of motion into the Kalb-Ramond equations of motion and vice versa
\begin{eqnarray} \label{eq:deltaEOMs}
 \delta_\beta \tilde{\cal E}_{a b} = -4 \beta_{c(a} \tilde{\cal B}_{b)}{}^{c} \; , \;\;\;\;\;\; \delta_\beta \tilde{\cal B}_{a b} = \beta_{c[a} \tilde{\cal E}_{b]}{}^{ c}\; .
\end{eqnarray}
It coincides exactly with the transformation rules found in \cite{Baron:2023qkx}, corresponding to the case where all p-form potentials have been truncated. To see this, we first decompose the $e.o.m.$ of the NS-NS sector as 
\begin{eqnarray}
\tilde {\cal E}_{ab} = e^{-2d} {\cal E}_{ab} + \Delta {\cal E}_{ab}\; , \;\;\;\;\;\;\;\;\;\;\;\;\;\;\;
 \tilde {\cal B}_{ab} = e^{-2d} {\cal B}_{ab} + \Delta {\cal B}_{ab}\; .
\end{eqnarray}
Here $\Delta{\cal E}_{ab}$ and $\Delta{\cal B}_{ab}$ denote the RR contribution, ${\cal E}_{ab}$ and ${\cal B}_{ab}$ already satisfy an equation like (\ref{eq:deltaEOMs}), as was discussed in section 4.4 of \cite{Baron:2023qkx} and $e^{-2d}=\sqrt{-g} e^{-2\phi}$ is $\beta$ invariant. So, in order to prove (\ref{eq:deltaEOMs}) we have to verify that $\Delta{\cal E}_{ab}$ and $\Delta{\cal B}_{ab}$ satisfy the same equation.

 The first term in $\Delta\mathcal{E}_{ab}$ is proportional to the R-R sector of the Lagrangian, which has already been proven to be invariant under $\beta$ transformations (cf. section \ref{sec4.1}). Also the proportionality factor, the flat metric $g_{ab}$, is $\beta$-invariant. The transformation rule for the field strengths with planar indices reads
\begin{eqnarray}
        \delta_\beta F^{(p)}_{c_1\dots c_p} &=& e^{\mu_1}_{c_1}\dots e^{\mu_p}_{c_p}\delta_\beta F^{(p)}_{\mu_1\dots \mu_p} + p (\delta e^{\mu_1}_{[c_1})F^{(p)}_{|\mu_1|c_2\dots c_p]} \cr
        &=& -\frac{1}{2} (i_\beta b) F^{(p)}_{c_1\dots c_p} + [\beta\wedge F^{(p-2)}]_{c_1\dots c_p} - \frac{1}{2}[i_{\beta}F^{(p+2)}]_{c_1\dots c_p} 
\end{eqnarray}
from which the variation of $\Delta \mathcal{E}_{ab}$ is
\begin{eqnarray}
    \delta_\beta (\Delta \mathcal{E}_{ab}) =
\sum_{p} \frac{1}{2(p-1)!}\left(\vphantom{\frac12} \delta_\beta(\sqrt{-g})F^{(p)}_{a c_1\dots c_{p-1}}F_{b}^{(p)\, c_1\dots c_{p-1}} + 2 \sqrt{-g}\delta_\beta F^{(p)}_{(a|c_1\dots c_{p-1}}F_{|b)}^{(p)\, c_1\dots c_{p-1}}\right)\cr
\end{eqnarray}
The first term of $\delta_\beta F^{(p)}$ will cancel out that of $\delta_\beta \sqrt{-g}$ (cf. Eq. (\ref{eq:deltaMetrica}) ). The remaining two terms do not cancel out completely in the sum over all forms as happened in the Lagrangian (cf. Eq. (\ref{eq:Cancelation}) ) as the forms are not fully contracted, but rather yield
\begin{eqnarray}
 \delta_\beta \Delta\mathcal{E}_{ab} = \sqrt{-g}\sum_p \frac{1}{(p-2)!}\beta_{(a|c_1}F^{(p-2)}_{c_2\dots c_{p-1}}F_{|b)}^{(p)\, c_1\dots c_{p-1}} = -4\beta_{c_1(a}\Delta{\mathcal{B}_{b)}}^{c_1}
\end{eqnarray}
thus proving the first of the relations in (\ref{eq:deltaEOMs}). A similar thing occurs with the variation of $\Delta\mathcal{B}_{ab}$: one of the terms of $\delta_{\beta} (F^{(p)})F^{(p-2)} + \delta_\beta (F^{(p-2)})F^{(p)}$ will cancel that of $\delta_\beta \sqrt{-g}$, while the remaining terms will yield the other relation in equation (\ref{eq:deltaEOMs}).

As discussed in section 3.2 of {\it loc.cit.} the transformations of the $e.o.m.$ play an important role on the description of deformations of $\beta$-transformations leaving invariant the action, but not the Lagrangian. In addition, a covariant transformation rule at the level of $e.o.m.$ guarantees $\beta$ symmetry transforms solutions into new solutions and so can be used as a solution generating technique in 10d.

Regarding the equations of motion of the p-forms, we start by noticing that 
\begin{eqnarray}
\delta_{\beta}[ d(e^{-b}\wedge * F)]&=&
d[  \delta_{\beta}(e^{-b})\wedge * F + e^{-b}\wedge \delta_{\beta}(* F) ]\cr
&=& d\left[\vphantom{\frac12}\right.(\beta+ [b\beta b])\wedge e^{-b}\wedge *F \cr 
&&
\;\;+\; e^{-b}\wedge \left(-\frac12 (i_{\beta}b)*F +\frac12 i_\beta(*F) -[b\beta*F]-\beta\wedge *F\right)
\left.\vphantom{\frac12}\right]\cr
&=& \frac12 d\left[(i_{\beta} (e^{-b})\wedge *F + e^{-b}\wedge i_{\beta}(*F) - 2 e^{-b}\wedge [b\beta*F]\right]\cr
&=&
\frac12 d[ i_{\beta}(e^{-b}\wedge * F)] \;. \label{vareomAux}
\end{eqnarray}
In the third equality we have used identity (\ref{ibetaeb}) with $b\to-b$, in the last line we used instead (\ref{ibetaAB}). 
Hence, we obtain 
\begin{eqnarray}
\delta_{\beta}[ d(e^{-b}\wedge * F)]
=\frac12  i_{\beta}[ d(e^{-b}\wedge * F)] \;.\label{eomFAux}  
\end{eqnarray}
This readily follows from (\ref{vareomAux}) and the identity
\begin{eqnarray}
d\left(i_{\beta}A^{(q)}\right) = i_{\beta}\left(dA^{(q)}\right)\;,       
\end{eqnarray}
which is valid for any $q$-form $A^{(q)}$ and is a consequence of (\ref{betaConstraint}) and the constancy of $\beta^{\mu\nu}$.

With (\ref{eomFAux}) at hand we can then proceed by following similar steps than those of section \ref{sec4.3} to obtain $\delta_\beta \left[*\left(d\left[e^{-b}\wedge *F\right]\right)\right]$. We find 
\begin{eqnarray}
\delta \tilde{\cal E}^{(p)}_{a_1 \dots a_p} =  (\tilde{\cal E}^{(p-2)}\wedge \beta)_{a_1 \dots a_p} - [\beta b \tilde{\cal E}^{(p)} ]_{a_1 \dots a_p} \;,  
\end{eqnarray}
a formal expression where it is implicitly assumed that $\tilde{\cal E}^{(p-2)}$ vanishes for $p<2$. Here $[\beta b \tilde{\cal E}^{(p)} ]$ is defined in analogous fashion as (\ref{AbetaB}), 
\begin{eqnarray}
    [\beta b \tilde {\cal E}^{(p)}] = \frac{1}{(p-1)!}\beta_{a_1 c}\; b^{c d} \;\tilde {\cal E}^{(p)}_{d a_2\dots a_{p}} e^{a_1}\wedge \dots\wedge e^{a_p} \;,\;\;\;\;\;\; e^{a}=e_{\mu}{}^{a} dx^{\mu}\;.
\end{eqnarray}

\subsection{Closure}\label{sec4.5}
We close the analysis of consistency of the $\beta$-transformations as an effective symmetry in 10 dimensions by verifying closure of the algebra. The local transformations of the NS-NS sector combined with $\beta$ symmetry and its associated bracket was studied in \cite{Baron:2022but} and \cite{{Baron:2023qkx}}. Here we extend it by considering also the local transformations in the R-R sector
\begin{eqnarray}
  \delta_{\xi} D^{(p)}_{\mu_1 \dots \mu_p} &=& \xi^{\rho}\partial_{\rho} D^{(p)}_{\mu_1 \dots \mu_p} + p\partial_{[\mu_1}\xi^{\rho} D^{(p)}_{|\rho|\mu_2 \dots \mu_p]}  \cr
    \delta_{\Lambda} D^{(p)}_{\mu_1 \dots \mu_p} &=& 0\;, \;\cr
  \delta_{\lambda}D^{(p)}_{\mu_1 \dots \mu_{p}} &=& p(p-1) \partial_{[\mu_1}\lambda_{\mu_2} D^{(p-2)}_{\mu_3 \dots \mu_p]}\cr
  \delta_{\chi} D^{(p)}_{\mu_1 \dots \mu_p} &=& p \partial_{[\mu_1}\chi^{(p-1)}_{\mu_2 \dots \mu_p]}.
\end{eqnarray}
where $\delta_{\xi}\; , \delta_{\Lambda},\; \delta_{\lambda},\; \delta_{\chi}$ stand for diffeomorphisms, Lorentz, gauge transformations of the Kalb-Ramond and p-form potentials, respectively.  
Indeed, one finds that the condition (\ref{betaConstraint}) leads to the following bracket 
\begin{eqnarray}
[\delta_1,\delta_2]=-\delta_{12}\;,    
\end{eqnarray}
with 
\begin{eqnarray}\label{Closure}
\Lambda_{12}{}_{a b} &=& 2 \, \beta_{\underline{1}[a}{}^{c} \beta{\vphantom{\beta^{a}}}_{\underline{2}\, b] c} 
    + 2\, \Lambda_{\underline{1} [a}{}^{c} \Lambda{\vphantom{\Lambda^{a}}}_{\underline{2}\, b] c}
    + 2\, \xi_{[\underline{1}}^{\mu}\partial_{\mu}\Lambda{\vphantom{\xi^{\mu}}}_{\underline{2}] a b} \cr
\beta_{12} &=& 0\;,\cr
\xi_{12}^{\mu} &=& 2 \xi^{\nu}_{[\underline{1}} \partial_{\nu}\xi_{\underline{2}]}^{\mu} + \beta_{[\underline{1}}^{\mu\nu} \lambda{\vphantom{\xi^{\mu}}}_{\underline{2}] \nu} \;  ,\cr
\lambda_{12 \mu}&=&
4 \,\xi_{[\underline{1}}^{\nu} \partial_{[\nu}\lambda_{\underline{2}] \mu]} \cr
\chi^{(p)}_{12}{}_{\mu_1 \dots \mu_{p}} &=& -\beta^{\rho\sigma}_{[\underline{1}} \chi^{(p+2)}_{\underline{2}]\rho\sigma \mu_1 \dots \mu_{p}} + 2 p(p-1) \lambda{\vphantom{\chi^{\mu}}}_{[\underline{1}}{}_{\mu_1} \partial{\vphantom{\chi^{\mu}}}_{\mu_2} \chi^{(p-2)}_{\underline{2}]\; \mu_3 \dots \mu_{p}]} +2(p+1) \xi_{[\underline{1}}^{\rho} \partial{\vphantom{\chi^{\mu}}}_{[\rho} \chi^{(p)}_{\underline{2}]\;\mu_1 \dots \mu_{p]}}\; .\;\;\;\;\cr &&
\end{eqnarray}

This completes the studies of consistency for the $\beta$-symmetry in the bosonic sector of type II theories.

\section{Conclusions}
In a recent article \cite{Baron:2022but}, it was shown that $SO(d,d)$ symmetry of KK reductions in the NS-NS universal sector of supergravity admits an uplift in the standard 10d formulation through the inclusion of $\beta$-transformations (\ref{NSbeta}), provided that the condition (\ref{betaConstraint}) is satisfied. The main advantage of this approach is that it enables the consideration of constraints on the effective action due to T-duality within the standard scheme of supergravity, without the need of performing compactifications, Lorentz non-covariant or non-geometric field redefinitions or the introduction of additional coordinates.

 This work extends this symmetry to the RR sector of (massive) type IIA and type IIB. The transformations of the RR-potentials and curvatures are displayed in equations (\ref{dbetaD}) and (\ref{deltaFcomponents}), respectively. We explicitly verified the exact invariance of the Lagrangian (without neglecting any total derivative terms) in sections \ref{sec4.1} and \ref{sec4.2}. Additionally, we conducted various consistency studies in sections \ref{sec4.3}, \ref{sec4.4} and \ref{sec4.5}. 

The invariance under $\beta$ transformations in the R-R sector can also be derived from a consistent truncation of the DFT extention of type II supergravities \cite{DFTtypeII}. In this formulation the RR potentials fit into Spin(10,10) multiplets $D$ by extending (\ref{Cspinor}) to 10 dimensions (see also \cite{GGtypeII} for a similar discussion in the context of Generalized Geometry). The kinetic terms for RR potentials are expressed in terms of a duality covariant extension of the exterior derivative, realized in terms of an $O(10,10)$ Dirac derivative $\Gamma^{M}\partial_M= \Gamma^{\mu}\partial_{\mu}+ \Gamma_{\mu}\partial^{\mu}$, where $\partial^{\mu}$ denotes differentiation with respect to the extra coordinates of DFT, duals to winding modes. 

The metric contracting the curvatures in (\ref{LR}) and the Kalb Ramond in the definition of the field strengths in (\ref{F}) are combined into a $Spin^{-}(10,10)$ object, so that T-duality invariance of the action and consistency relations ($e.g.$ with duality relations and $e.o.m.$ ) are strongly simplified. Explicitly breaking the $O(10,10)$ symmetry by choosing the supergravity section  $\partial_{M}=(\partial_{\mu},\partial^{\mu}) \to (\partial_{\mu},0)$, but still keeping the duality multiplets, leads to type II supergravity in a DFT scheme. Although this chosen section breaks the duality invariance, it is expect that the constraint (\ref{betaConstraint}) effectively restores it. Indeed, while this section is preserved by $Gl(d)$ and b-shift, a $\beta$-transformation  modifies it as $(\partial_{\mu},0)\to_{\beta}(\partial_{\mu},\beta^{\mu\nu} \partial_{\nu})$.  

In conclusion, it is expected that $\beta$-invariance is easily verified in this scheme. However, the price to pay is that, contrary to the scheme used in this paper, diffeomorphism and gauge invariance of the Kalb Ramond are no longer manifest.

There are some interesting extensions of this work, both in the NS-NS and the R-R sectors. Concerning the latter, it remains unclear whether there exists a consistent definition of $\beta$-symmetry within the standard formulation of type II theories, as there is no an off-shell mechanism to relate it to the democratic formulation considered in this paper. 

Regarding the NS-NS case, it is worth to investigate higher derivative deformations. It was argued in \cite{Hronek:2020xxi} that there exists a universal sector of bosonic, heterotic and type II effective actions at eight order in derivatives (those proportional to $\zeta(3)$ ) that cannot be embedded in a Double Field Theory formulation\footnote{At least, by using the standard (fundamental) representation of $O(d,d)$ and ignoring Chern-Simons like interactions.}. The approach presented here is free from these obstructions, thus it is worthwhile to explore whether there is a $\beta$-deformation constraining these couplings. If so, it could shed light on a potential DFT-like deformation, circumventing the obstructions mentioned in the aforementioned work. First corrections to ${\cal L}_{NS}$ of type II theories are precisely $\sim{\cal O}(\alpha'{}^3)$, making it the simplest scenario to study this problem.
\bigskip

{\bf Acknowledgements:} We warmly thank C. N\'u\~nez and D. Marqu\'es for discussions and comments on the manuscript. Special thanks are extended to Yuho Sakatani for providing detailed comments regarding the connection between $\beta$-transformations and Yang Baxter deformations.
Support by Consejo Nacional de Investigaciones Cient\'ificas y T\'ecnicas (CONICET), through the Grant PIP-11220200100981CO, Universidad Nacional de La Plata (UNLP) and Universidad de Buenos Aires (UBA) are also gratefully acknowledged.

\end{document}